\documentclass[prl,aps,twocolumn,showpacs,amsmath,amssymb,superscriptaddress]{revtex4-1}

\usepackage{graphicx}
\usepackage{color}
\usepackage{dcolumn}
\usepackage{bm}
\usepackage{epsf}
\usepackage{amsmath,amstext,amssymb}
\usepackage{enumitem}
\usepackage{hyperref}
\usepackage[ansinew]{inputenc}
\usepackage{epstopdf}

\newcommand{\be}{\begin{equation}}
\newcommand{\ee}{\end{equation}}
\newcommand{\bea}{\begin{eqnarray}}
\newcommand{\eea}{\end{eqnarray}}

\begin{document}

\title{Single particle microscopy with nanometer resolution}

\author{Georg Jacob}
\email{georg.jacob@uni-mainz.de}
\affiliation{QUANTUM, Institut f\"ur Physik, Universit\"at Mainz, Staudingerweg 7, 55128 Mainz, Germany}

\author{Karin Groot-Berning}
\affiliation{QUANTUM, Institut f\"ur Physik, Universit\"at Mainz, Staudingerweg 7, 55128 Mainz, Germany}

\author{Sebastian Wolf}
\affiliation{QUANTUM, Institut f\"ur Physik, Universit\"at Mainz, Staudingerweg 7, 55128 Mainz, Germany}

\author{Stefan Ulm}
\affiliation{QUANTUM, Institut f\"ur Physik, Universit\"at Mainz, Staudingerweg 7, 55128 Mainz, Germany}

\author{Luc Couturier}
\altaffiliation{Present address: Physikalisches Institut, Universit\"at Heidelberg, Im Neuenheimer Feld 226, 69120 Heidelberg, Germany}

\author{Ulrich G. Poschinger}
\affiliation{QUANTUM, Institut f\"ur Physik, Universit\"at Mainz, Staudingerweg 7, 55128 Mainz, Germany}

\author{Ferdinand Schmidt-Kaler}
\affiliation{QUANTUM, Institut f\"ur Physik, Universit\"at Mainz, Staudingerweg 7, 55128 Mainz, Germany}

\author{Kilian Singer}
\affiliation{QUANTUM, Institut f\"ur Physik, Universit\"at Mainz, Staudingerweg 7, 55128 Mainz, Germany}

\date{\today}

\begin{abstract}
We experimentally demonstrate nanoscopic transmission microscopy relying on a deterministic single particle source. This increases the signal-to-noise ratio with respect to conventional microscopy methods, which employ Poissonian particle sources. We use laser-cooled ions extracted from a Paul trap, and demonstrate remote imaging of transmissive objects with a resolution of 8.6$\;\pm\;$2.0$\,$nm and a minimum two-sample deviation of the beam position of 1.5$\,$nm. Detector dark counts can be suppressed by 6 orders of magnitudes through gating by the extraction event. The deterministic nature of our source enables an information-gain driven approach to imaging. We demonstrate this by performing efficient beam characterization based on a Bayes experiment design method.
\end{abstract}

\pacs{37.10.Ty, 41.75.-i, 68.37.-d} 

\maketitle

The advantages of using single particles for sensing and imaging comprise state control down to the quantum regime, minimal sample disturbance, and well-defined couplings. This allows for nanoscopic sensing of optical, electric or magnetic fields. Various techniques have been developed to position and scan the location of such sensors in the nanometer range. Local sensing can be achieved with single ions or atoms inside a trap~\cite{guthohrlein2001single}, alternatively, a nano diamond containing a color center~\cite{degen2008scanning} or a single fluorescent molecule~\cite{michaelis2000optical} attached to a scanning tip of a near-field microscope can be used.

By contrast, statistical multi-particle irradiation in transmission microscopy~\cite{reimer2008transmission} allows for remote high resolution imaging of partially transmissive objects on the nanometer scale with increased depth of field. To reach high resolution, probe particles have to be tightly focused to interact with the sample, whose position is scanned with respect to the beam. Electrons or ions~\cite{ward2006helium} with high energy and flux are spatially filtered to reach nanometer resolution. Approaches employing neutral Helium atoms~\cite{koch2008imaging} or laser cooled Cesium ions from a cold magneto optical trap have achieved high brightness and nanometer spot sizes~\cite{knuffman2013cold}. Due to the statistical nature of these sources only shot-noise limited images can be obtained.

To combine the latter approach which offers remote sensing and high spatial resolution with the advantages of single particle sensing various schemes have been suggested. Proposals employ Rydberg atoms and the dipole blockade mechanism~\cite{ates2013fast,urban2009observation,miroshnychenko2009observation}, alkali atoms in magneto-optical traps~\cite{hill2003atoms}, or an atomic ensemble within an optical lattice during the transition to the single occupancy Mott insulator~\cite{greiner2002quantum}.

\begin{figure}
\begin{center}
\includegraphics[width=0.9\columnwidth]{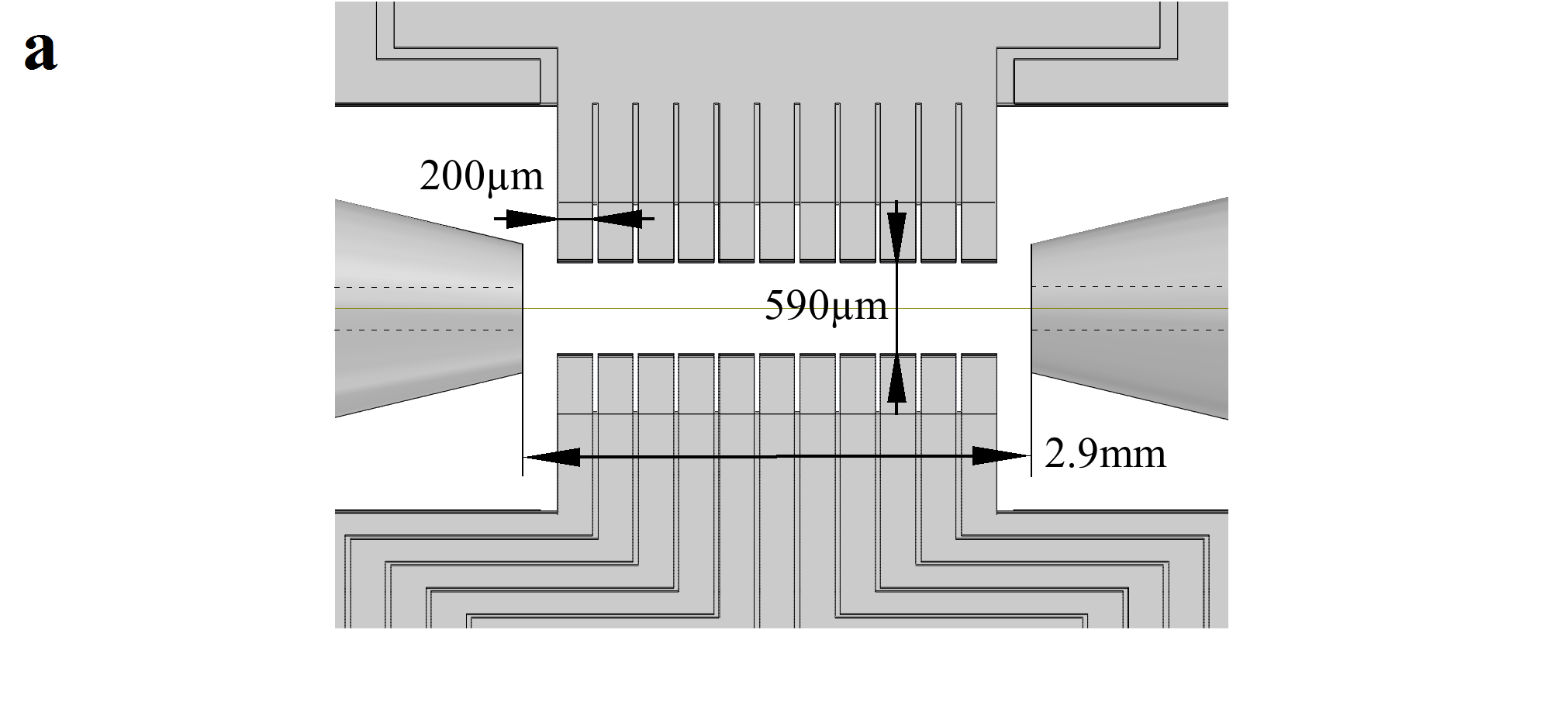}
\includegraphics[width=0.9\columnwidth]{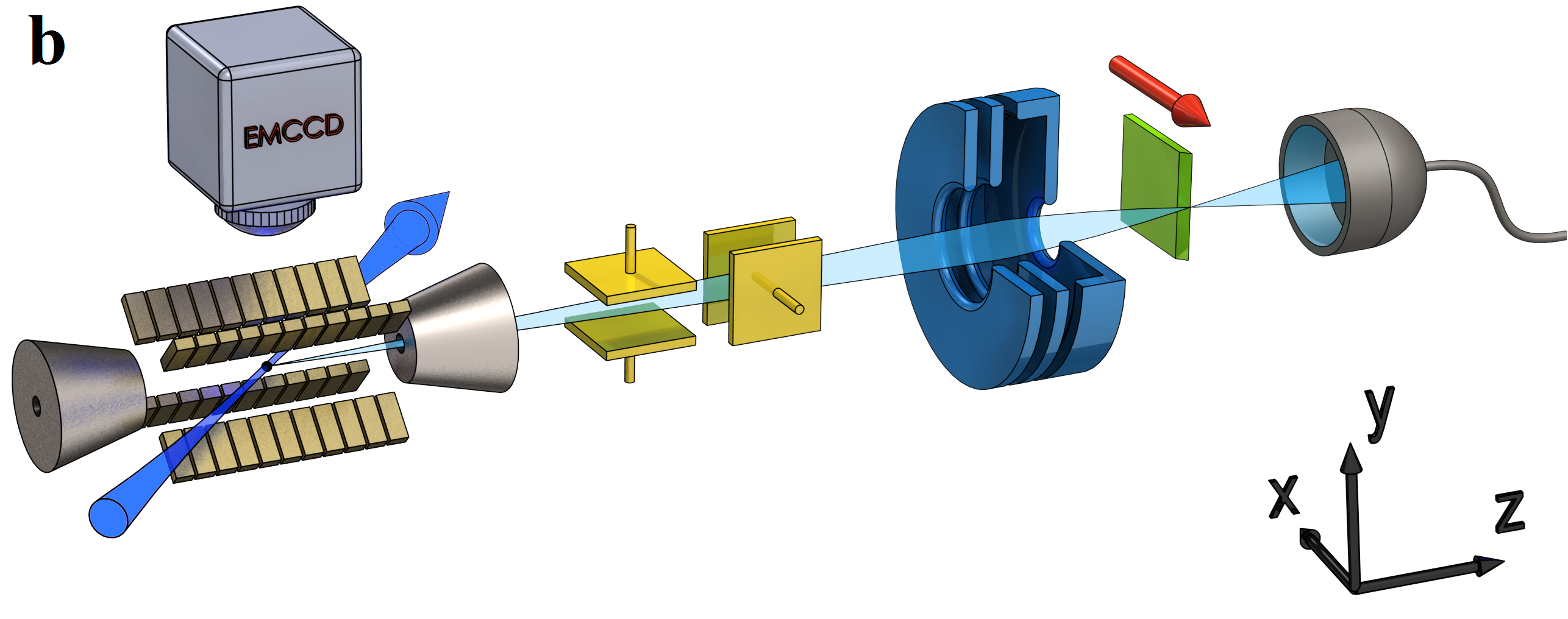}
\caption{(a) True to scale image of the trap geometry denoting important distances, as seen from the CCD. (b) Sketch of the single-ion microscope. The ion trap consists of segmented electrodes and end-caps. Laser-cooling (blue arrow) and imaging system are shown. Deflection electrodes (yellow) and einzel-lens (blue) are placed along the ion extraction pathway (light-blue). In the focal plane a profiling edge or alternatively a transmissive mask (green) is placed on a three-axis nano-translation stage (not shown). Single ion counting is performed with a secondary electron multiplier device.\label{fig:trap}}
\end{center}
\end{figure}

Our approach implements a transmission microscope based on a deterministic ion source delivering single laser-cooled $^{40}$Ca$^+$ ions from a linear segmented Paul trap~\cite{schnitzler2009deterministic,izawa2010controlled}. It is constructed from four micro-fabricated alumina chips arranged in an X-shaped configuration and two pierced metal end-caps (see Figure~\ref{fig:trap}a). Each chip comprises 11 electrodes for shaping the potential along the axial direction. We operate the device at trap frequencies $\omega/(2\pi)$ = 0.58$\,$MHz and 1.4$\,$MHz for the axial and radial directions. Calcium ions are loaded by photo-ionization and laser cooled on the S$_{1/2}$ to P$_{1/2}$ dipole transition. For high ion throughput, an automated loading of a predefined number of ions is implemented: Initially a random amount of ions are trapped, the ion number is counted by imaging the ion fluorescence on a CCD-camera and excess ions are removed by lowering the axial trap potential with a predefined voltage sequence. After automatic verification of the exact number of trapped ions, using laser light near 397$\,$nm under optimized Doppler-cooling conditions, the cold ions are extracted along the axial direction of the trap by application of an acceleration voltage of -2.5$\,$keV to one end-cap. It is controlled by a fast high voltage solid-state switch, whose jitter of less than 1$\,$ns allows for synchronization of the extraction time to the phase of the radio-frequency trap-drive ($\Omega/(2\pi)$ = 23$\,$MHz) with adjustable delay. With this method we attain rates for loading and extraction of single ions of 3$\,$Hz, corresponding to an average flux of about 0.5$\,$atto-ampere. Ions leave the trap passing through a 200$\,\mu$m diameter hole in the end-cap. For alignment and scanning of the beam, two pairs of deflection electrodes are placed along the ion trajectory. They are located 46$\,$mm and 67$\,$mm respectively from the center of the trap (see Figure~\ref{fig:trap}b). For focusing of the beam, a electrostatic einzel-lens is placed 332$\,$mm from the center of the trap. It consists of three concentric ring shaped electrodes with an open aperture of 4$\,$mm. The geometry parameters are optimized by electrostatic simulations~\cite{Singer:2010} to minimize spherical aberration. Chromatic aberration are highly suppressed due to the narrow velocity distribution the of ions. We measure a half width half maximum spread of $\Delta v=8\,$m/s at an typical average speed of $10^5\,$m/s. For imaging a partially transmissive object is placed on a three-axis translation stage. This can be either a nano-structured test sample or a profiling knife-edge. The transmitted ions are detected by a secondary electron multiplier. Image information is generated by recording transmission events for a well defined number of extractions while scanning the object position in the focal plane.

\begin{figure}
\begin{center}
\includegraphics[width=0.9\columnwidth]{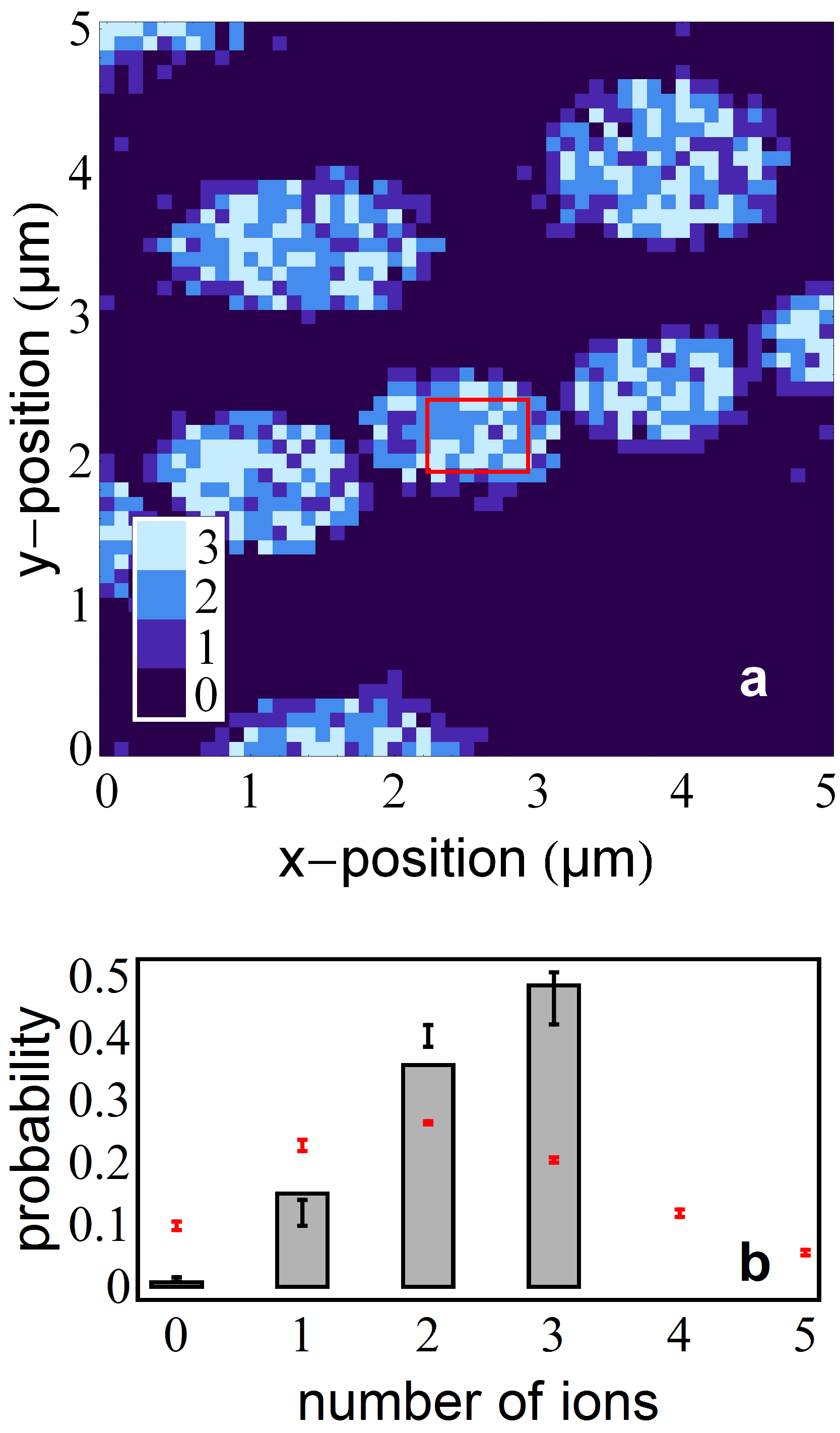}
\end{center}
\caption{Imaging a transmissive sample. The Quantifoil Multi A carbon film is scanned using three ions at each lateral position. (a) The image is taken with a resolution of (100x100)$\,$nm$^2$ per pixel, (b): counting statistics of the region in (a) within the red box. A strongly sub-Poissonian distribution (gray bars) is observed with a mean of 2.33$\;\pm\;$0.07 and a Mandel Q-parameter of -0.76. For comparison a Poissonian (red error bars) and a Binomial distribution (black error bars) with identical means are depicted. The error bars are calculated for the two distributions by using the standard error of the mean of the measured number of ions.
\label{fig:scanplot}}
\end{figure}

We demonstrate the imaging of transmissive structures by scanning a carbon film~\footnote{QUANTIFOIL Multi A, http://www.quantifoil.com} with defined hole structures that is commercially available for electron microscopy testing and adjustment (see Figure~\ref{fig:scanplot}a). The small holes are specified to have a diameter of 1$\,\mu$m. We find that the hole dimensions in vertical and horizontal direction are 900$\,$nm and 1400$\,$nm, respectively. Here, the resolution is limited to 100$\,$nm by the used pixel size. A high contrast image is obtained with only 3 consecutive ion extractions per pixel: the entire information in the picture is based on 1,515 transmission events out of 7,803 extracted ions. The average count number for completely transmissive parts of the sample is 2.33$\;\pm\;$0.07, consistent with a detector efficiency of 77$\;\pm\;$2\%. A histogram (grey bars in Figure~\ref{fig:scanplot}b) of the counts at each pixel within the marked area~\footnote{Here a higher pixel density of (50x50)$\,$nm$^2$ is used.} (red box in Figure~\ref{fig:scanplot}a) fits to a binomial distribution (black error bars). Such counting statistics are fundamentally different to that of a probabilistic Poissonian source with the same mean counts per pixel (red error bars). Especially for low mean count numbers, the sub-Poissonian character with a measured Mandel Q-parameter~\cite{mandel1979sub} of -0.76 provides considerably better image contrast~\cite{delaubert2008quantum} as compared to Poissonian sources with Q$\;=\;$0. Additionally, the deterministic source allows for gating of the detector (typically less than 200$\,$ns), such that its dark count rate ($<\;$100 $s^{-1}$) does not affect the image contrast. Consequently, the observed image does not show any noise in the pixels corresponding to the non-transmissive parts of the object.

\begin{figure}
\begin{center}
\includegraphics[width=0.9\columnwidth]{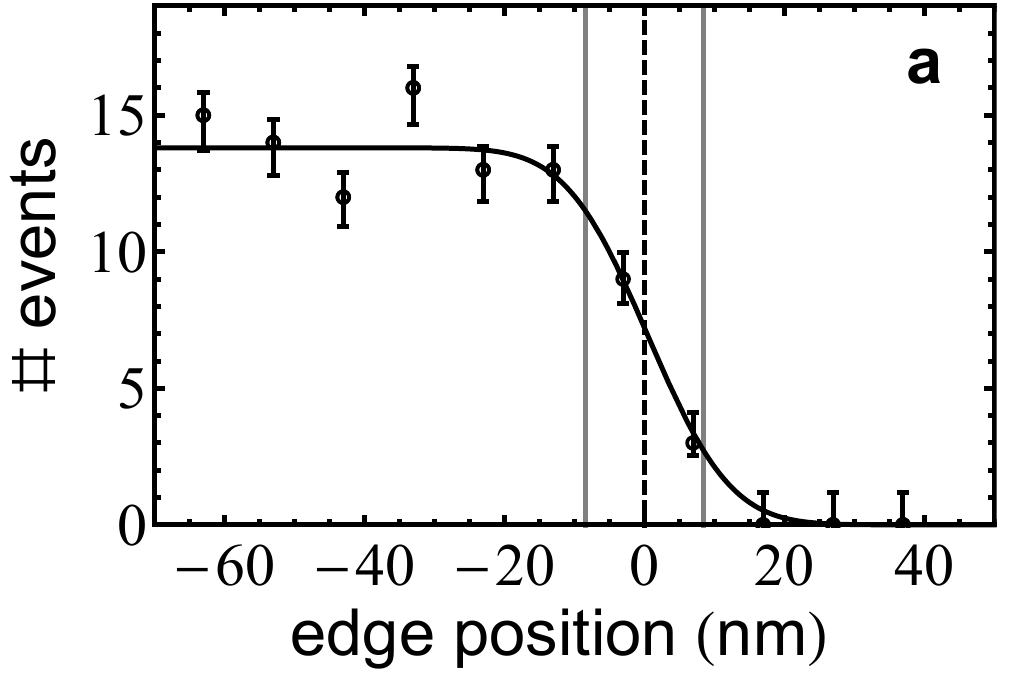}
\end{center}
\begin{center}
\vspace{-7mm}
\end{center}
\begin{center}
\includegraphics[width=0.9\columnwidth]{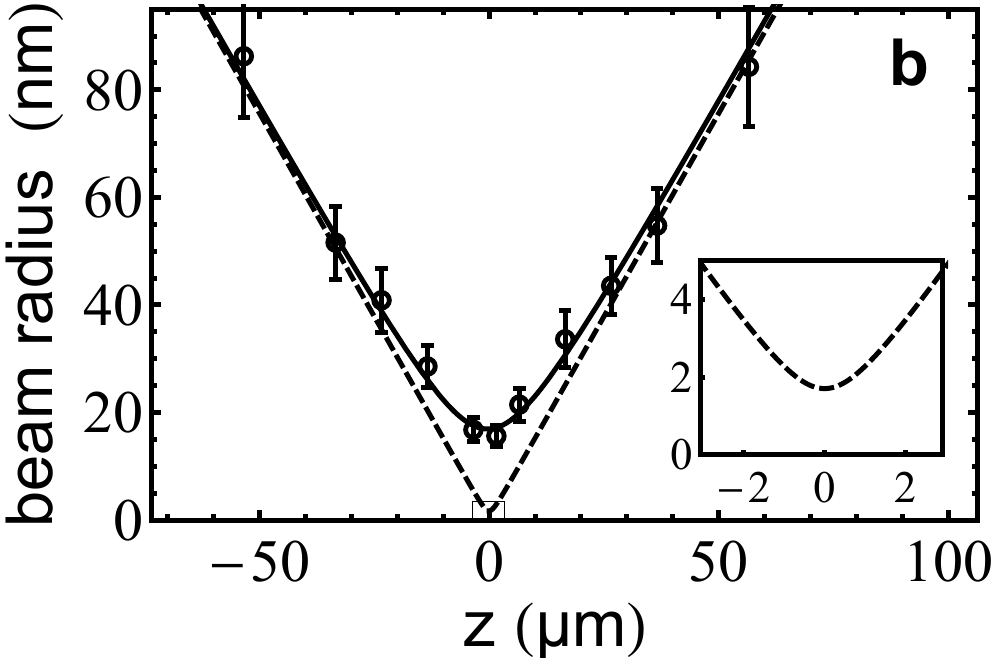}
\end{center}
\caption{Beam profiling measurements. (a) the circles represent the number of detected ions under optimal conditions from 20 single ion extractions at the corresponding profiling edge position, in total 220 ion shots with 4 minutes data acquisition time.
Error bars depict the standard deviation from a binomial distribution. The dashed black line illustrates the center position and the grey lines the 1-sigma radius of the beam. The continuous black line fits a Gaussian error-function $p(x)=\frac a2\left[1+\textrm{erf}\left(\frac{x-x_0}{\sigma\sqrt 2}\right)\right]$ to the data with $a=\;$14.1$\;\pm\;$0.5 and $\sigma=\;$8.6$\;\pm\;$2.0$\,$nm. (b) The 1-sigma beam radius (black circles) dependence along the beam axis, near a 11$\,$mm distance from the lens. The data (in total $\sim\;$2$\,$h data acquisition, about 5,000 ions) is fitted (solid line) yielding a source size of about 500$\,$nm. The error bars are obtained from the standard deviation of the probability density function of the Bayesian experimental design method. The dashed line shows the same function assuming a source size of 50$\,$nm. A zoomed-in plot can be seen in the inset.
\label{fig:focus}}
\end{figure}

For an accurate determination of the beam characteristics and the spatial resolution, we perform a beam profiling in the focal plane (Figure~\ref{fig:focus}a). At each position of the profiling edge, 20 ions are extracted and the number of detector events are recorded. To obtain the beam parameters, the data is fitted with a Gaussian error function. Under optimal conditions this yields 8.6$\;\pm\;$2.0$\,$nm for the 1-sigma radius of the beam. In order to determine the waist function, profiling measurements are performed along the $z$-axis of the beam (Figure~\ref{fig:focus}b). For this purpose  the deterministic nature of our source allows for maximizing the information gain per probe event by using the Bayesian experimental design technique~\cite{lindley1956measure,guerlin2007progressive,pezze2007phase,brakhane2012bayesian}. With this method quantities such as the beam radius, and position can be obtained more efficiently. This is achieved by choosing experimental parameters such that the expected information gain is maximized before the next ion is extracted. In our case the experimental parameter $\xi$ (often referred to as design) is the profiling edge position for the next measurement. In order to find the next optimal position of the profiling edge, we define a utility $U(y,\xi)$ as the difference in Shannon entropy between $p(\theta)$ the prior and  $p(\theta|y,\xi)$ the posterior probability density function (PDF) of $\theta$ the parameters to be measured (beam position, radius and detector efficiency). It is calculated as a function of all possible edge positions $\xi$ and measurement outcomes $y\in \lbrace 0,1 \rbrace$: 
\begin{equation*}
U(y,\xi)= \int \ln(p(\theta|y,\xi))p(\theta|y,\xi) d\theta-\ln(p(\theta))p(\theta) d\theta,
\end{equation*}
where the posterior PDF is obtained using the Bayesian update rule
\begin{equation*}
p(\theta|y,\xi)=\frac{p(y|\theta,\xi)p(\theta)}{p(y,\xi)}.
\end{equation*}
The expected utility independent of the observation is
\begin{equation*}
U(\xi)=\int U(y,\xi)p(y,\xi) dy=\sum_{y=0}^1 U(y,\xi)p(y,\xi).
\end{equation*}
Finally, the profiling edge position yielding the highest utility and therefore the maximal difference in entropy is chosen, guaranteeing maximal information gain with the next extraction of a single ion. Using the outcome of the actually conducted experiment, the Bayesian method additionally allows for iteratively updating the prior PDF of the parameters to be determined. The procedure is then repeated until a accuracy goal is reached. We fit the waist of the ion beam by the function
\begin{eqnarray}
w(z)=\sqrt{\frac{(z-f)^2 \left(s^2 \alpha ^2+S^2\right)}{f^{2}}+\frac{2s \alpha ^2 z(f-z)}{f}+\alpha ^2 z^2},\nonumber
\end{eqnarray}
with $z$ denoting the position along the extraction axis, $f=11.86\,$mm the focal length, $\alpha=51.7\;\pm\;$1.2$\,\mu$rad the beam divergence, $s=332\,$mm the distance between source and lens and $S=496\;\pm\;$30$\,$nm the source size. The model does not take lens errors into account, as the contribution to the minimal focal radius amounts to $r_{\mathrm{min}}=c_2r_\mathrm{inc}^3/4f=\;$0.2$\,$nm~\cite{egerton2005physical}. Here $r_\mathrm{inc}=17.17\,\mu$m is the radius of the incident beam, $f=11.86\,$mm is the focal length and $c_2=\;$2.0$\;\pm\;$0.2$\,$mm$^{-1}$ is the spherical aberration-coefficient of the einzel-lens. With the deflection electrodes we carefully adjust the incident ion beam position on the center of the electrostatic lens. From the measured waist function (see Figure~\ref{fig:focus}b) we find a focal radius of 16.9$\;\pm\;$1.0$\,$nm. 
The beam divergence is fully consistent with a value of 60$\,\mu$rad divergence from the ion trap as measured independently without the focusing lens. The fit to the model allows for extracting the effective source radius $S$. The a value of about 500$\,$nm, it is well above the single ion wave packet size of 50$\,$nm (pertaining to a radial trap frequency of 1.4$\,$MHz at a temperature of 2$\,$mK). This discrepancy is attributed to mechanical vibrations and thermal drifts of the apparatus. The total length between trap chamber and detector is about 0.3$\,$m and the setup is placed on a regular air-damped optical table. Also relevant for effective source point fluctuations could be residual electrical noise on the deflection electrodes and drifts of the optimal voltage setting for micro-motion compensation. Due to the frequent photoionization loading cycles and the applied high voltage pulses on the extraction end-cap, we find typical variations of micro-motion compensating voltages of 3$\,\%$. Applying the model function for the waist size and assuming a 50$\,$nm source radius, which corresponds to a Doppler cooled ion without any additional source point fluctuations, results in a calculated focal radius of 1.7$\,$nm over a $\sim2\,\mu$m depth of field (dashed line in Figure~\ref{fig:focus}b). This result is supported by classical numerical trajectory Monte-Carlo simulations~\cite{Singer:2010} predicting the same focal radius. For ions cooled to the motional ground state with a wave packet size of about 15$\,$nm, a further reduction of the focal radius may be achieved. A resolution range well below 1$\,$nm is attained by transmission electron and He ion microscopes~\cite{ward2006helium}.

\begin{figure}
\begin{center}
\includegraphics[width=0.9\columnwidth]{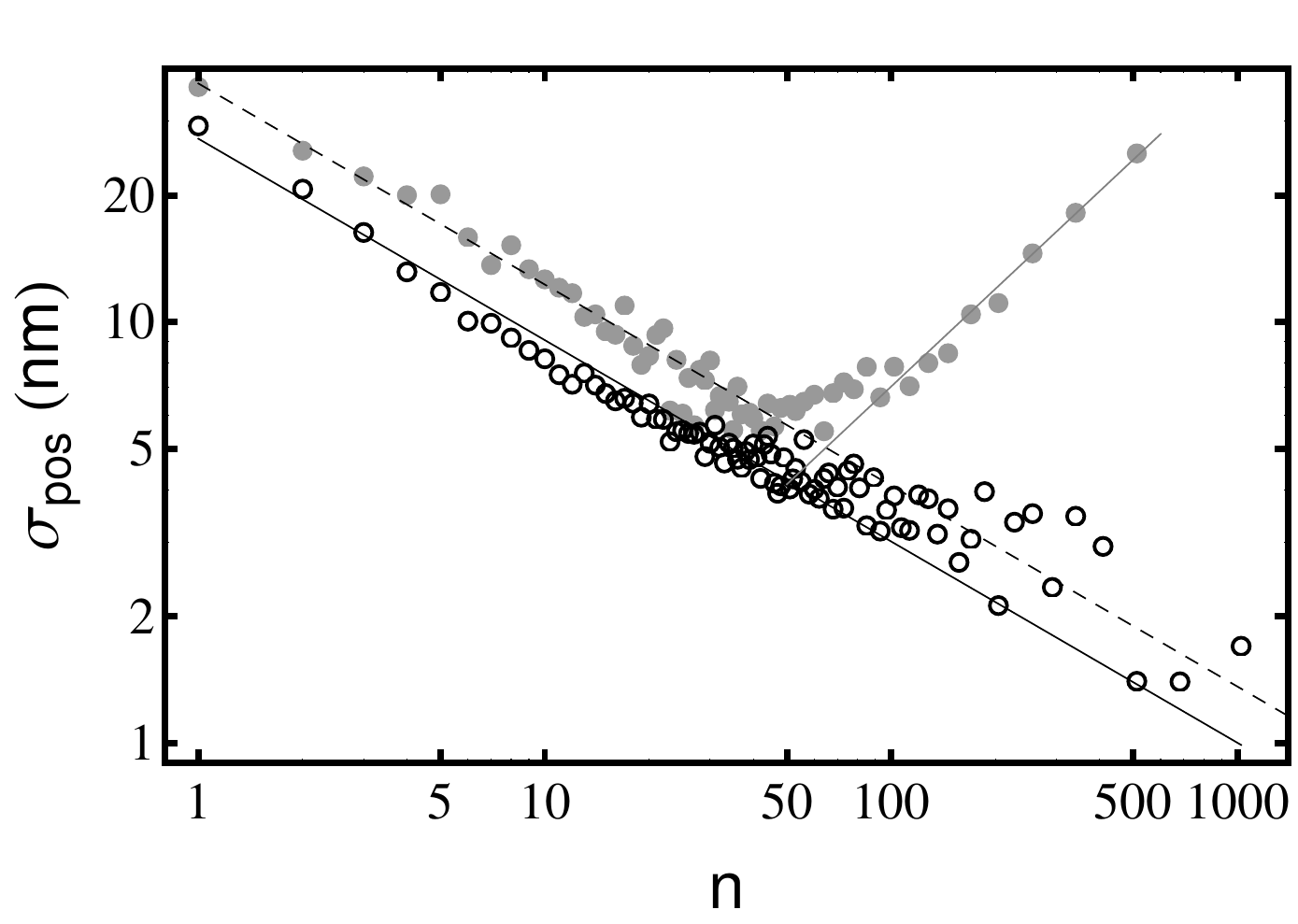}
\end{center}
\caption{Log-log plot of the two-sample deviation of the beam position. Measurement data obtained under best conditions (black circles) in contrast to the same measurement under less stable conditions (grey, see text). The fit (black solid line) to the beam position deviations (black circles) show a slope of -0.48$\;\pm\;$0.01 . \label{fig:two-sample}}
\end{figure}

The long-term stability of the apparatus is an essential prerequisite for high resolution imaging of large samples. Similar to the Allan-deviation technique, pioneered in the field of shot-noise limited atomic clocks~\cite{Santarelli1999QuantumProjectionNoise}, we have evaluated the two-sample deviation of the lateral position of the beam (see Figure~\ref{fig:two-sample}). To this goal 2,048 profiling edge measurements (similar to those in Figure~\ref{fig:focus}a) are carried out repetitively. Every measurement comprises 26 different edge positions separated by 10$\,$nm, each probed with a single ion. In total the data set contains 53,248 ion extraction events and is recorded during a non-interrupted acquisition time of 18 hours. For the data analysis, the counts per edge position of $n$ consecutive profiling measurements~\footnote{The number $n$ of ions per edge position is ranging between 1 and 1024.} are summed up and fitted to extract beam position $\bar{x}(n)$. The $n$-th two-sample deviation (see Figure~\ref{fig:two-sample}) is calculated with 

\begin{equation*}
\sigma_{\text{pos}}(n) =\sqrt{ \frac{1}{2}\langle\left( \bar{x}_{i+1}(n)-\bar{x}_i(n)\right)^2\rangle}.
\end{equation*}

If the beam pointing fluctuations are dominated by statistical noise, the value of $\sigma_{\text{pos}}$  is expected to improve with $\sqrt{n}$, showing a falling slope of 1/2 in a log-log plot, which is in full agreement with our data. 
Under stable thermal and constant trapping conditions~\footnote{We continuously heat the calcium oven and avoid the use of a titanium sublimation pump. The radio frequency-drive and all lasers are continuously switched on.}, the minimal two-sample deviation of the beam position yields a long-term beam-pointing stability of 1.5$\,$nm (Figure~\ref{fig:two-sample}, black circles). We recorded the same data without these stabilization measures to demonstrate their effectiveness (Figure~\ref{fig:two-sample} grey dots) where after about $n=\;$40 measurement sequences corresponding to 20$\,$min, we observe the onset of drifts in the position of the beam. 

Due to the unique statistical properties of this system, the deterministic transmissive single-ion microscopy is a minimal disturbing imaging technique and could resolve features which would be unobservable with conventional techniques. It is applicable to insulating substrates due to the prevention of surface charging. The triggered non-Poisson source allows for high contrast and efficient suppression of dark counts. Moreover, the time of flight information could be used \textit{e.g.} by switching the focusing fields and thus circumventing the resolution-limiting Scherzer-Theorem~\cite{schonhense2002correction}, which states that a rotationally symmetric ion optical lens with static electromagnetic fields under the absence of space charges always exhibits unavoidable spherical and chromatic aberrations. The single-ion microscope offers versatile applications, as it can be used with a large variety of atomic and molecular ions that can be cooled sympathetically~\cite{drewsen2004nondestructive}. We have verified efficient loading and cooling of N$_2^+$ ions and CaO$^+$ ions. This paves the way for future applications and precision experiments: The apparatus could be used for the deterministic fabrication of solid state quantum devices such as coupled nitrogen vacancy color centres~\cite{dolde2013room}. Single phosphorous nuclear spins in silicon~\cite{kane1998silicon,jamieson2005controlled,pla2013high}, cerium or praseodymium in yttrium orthosilicate~\cite{kolesov2012optical} might also be implanted on a nano-scale and with respect to markers on the sample. Furthermore, through optical pumping one may implement a fully spin-polarized source for sensing magnetic polarization of surfaces as in electron microscopy~\cite{duden1998spin}. The combination of control of the internal and external degrees of freedom of the ion would allow for the realization of matter wave interferometry with single ions~\cite{arndt2012focus,hasselbach2010progress}.

\paragraph*{Acknowledgments}The authors acknowledge discussions with S. Prawer, G. Sch\"onhense, S. Dawkins, R. Gerritsma and C. Schmiegelow. The team acknowledges financial support by the Volkswagen-Stiftung, the DFG-Forschergruppe (FOR 1493) and the EU-projects DIAMANT and SIQS (both FP7-ICT).

\bibliography{impbib}

\end{document}